\documentclass{article}
\usepackage[english]{babel}
\usepackage[letterpaper,top=2cm,bottom=2cm,left=3cm,right=3cm,marginparwidth=1.75cm]{geometry}
\usepackage{multicol}
\usepackage{multirow}
\usepackage{amssymb}
\usepackage{amsmath}
\usepackage{booktabs}
\usepackage{makecell}
\usepackage[table]{xcolor}
\usepackage{tcolorbox}
\tcbuselibrary{skins}
\usepackage{graphicx}
\usepackage{subcaption}
\usepackage{xspace}
\usepackage{wrapfig}
\usepackage{enumitem}
\usepackage{pifont}
\usepackage{tikz}
\usepackage{pgfplots}
\pgfplotsset{compat=1.18}
\usepackage{xurl}
\usepackage[colorlinks=true,allcolors=blue]{hyperref}
\newcommand{\tech}{\textrm{\mdseries\textsc{SkillMorph}}\xspace}
\newcommand{\keywords}[1]{\noindent\textbf{Keywords:} #1}

\title{Trajectory-Guided Fault Localization for Agent Skill Evolution}
\author{
Yu Ge\textsuperscript{1},
Linna Xie\textsuperscript{1},
Zhong Li\textsuperscript{1}\thanks{Corresponding author.},
Yu Pei\textsuperscript{2},
Tian Zhang\textsuperscript{1}\thanks{
Emails for all authors:
yuge@smail.nju.edu.cn,
xieln@smail.nju.edu.cn,
lizhong@nju.edu.cn,
yupei@polyu.edu.hk,
ztluck@nju.edu.cn.
}\\[0.5em]
\textsuperscript{1} Nanjing University\\
\textsuperscript{2} The Hong Kong Polytechnic University
}
\date{}
\hypersetup{
  pdftitle={Trajectory-Guided Fault Localization for Agent Skill Evolution},
  pdfauthor={Yu Ge, Linna Xie, Zhong Li, Yu Pei, Tian Zhang},
  pdfsubject={Trajectory-guided fault localization for agent skill evolution},
  pdfkeywords={Agent Skills, Skill Evolution, Fault Localization, Trajectory Analysis}
}

\begin{document}
\maketitle
\begin{abstract}
Agent skills provide reusable guidance for code agents, but incomplete or unsuitable guidance can impair task execution.
To reduce the manual effort of skill refinement, recent approaches use LLMs to generate revisions from execution feedback.
However, grounding these revisions in explicit behavioral evidence remains challenging.
To address this gap, we propose \tech, a skill-evolution approach based on trajectory-guided fault localization in agent skills.
Its core idea is to link execution evidence to specific skill contents before generating revisions.
Specifically, \tech compares failure and success evidence in abstracted trajectories across repeated runs and tasks, incorporating changes between evolution loops to identify suspicious actions.
It then uses these suspicious actions to localize edit sites in the skills and generate corresponding revisions.
Experiments on SWE-Skills-Bench and CannBot show that the skills evolved by \tech consistently achieve higher trial-level accuracy and execution consistency than the original skills and those from four existing skill-evolution methods.
We have also applied \tech to automated kernel generation with an AI operator-development team, which has accepted 6 skill-revision pull requests.
\end{abstract}

\section{Introduction}

Agent Skills have become a common paradigm for supporting large language model (LLM)-based agents in domain-specific tasks such as software engineering~\cite{swe-skills-bench}, data analysis~\cite{huang2026skills-data-analysis}, and document processing~\cite{anthropic-agent-skills}.
Specifically, a skill packages reusable instructions, scripts, and reference materials for a specialized workflow.
When carrying out a task, the agent loads the relevant skills and takes actions under their guidance, leveraging the experience encoded in those skills to solve the task more effectively.
In practice, however, incomplete guidance or unsuitable workflows can lead the agent to take unsuitable actions, which limits skill effectiveness or even impairs task execution~\cite{skillaxe,swe-skills-bench}.
Addressing these deficiencies generally relies on domain experts to inspect the agent's actions, trace the problematic ones back to the skill content that guided them, and iteratively revise and evaluate the guidance~\cite{evoskill,anthropic-agent-skills}.
This manual process makes skill refinement costly and time-consuming, and difficult to scale as task and skill collections grow.

To reduce manual effort, several approaches have recently been proposed for \textbf{\textit{skill evolution}}, which aims to improve agent performance by automatically refining skills based on execution feedback~\cite{skillopt,skilladaptor,evoskill,skillgrad,skillrevise,skillscoach,skillevolver}. 
Most of these techniques follow an iterative process with two steps: (1) executing a set of evolution tasks using the current skill set to collect execution feedback, and (2) prompting an LLM with this feedback to produce a revision of the skill content.
Despite the effectiveness of existing skill-evolution approaches, three challenges remain.

\emph{First}, it is challenging to ground skill revision in explicit behavioral evidence.
Existing approaches generally prompt an LLM to generate revisions directly from execution feedback, such as task outcomes, case-level scores, or whole trajectories~\cite{skillopt,skilladam,evoskill,skillaxe,coevoskill}.
Consequently, which behaviors are problematic is implicitly determined by the LLM as it reads the feedback and writes the revision, leaving no record for later inspection or comparison.
This results in two consequences.
On the one hand, without an explicit link between problematic behaviors and the skill content that needs correction, the revisions are likely to alter content that merely appears relevant to the observed problem while leaving the content requiring correction unchanged.
Recently, ContractSkill~\cite{contractskill} achieved explicit localization by representing a skill as an ordered sequence of steps with postconditions and using a deterministic verifier to identify the failing step.
However, it assumes that the skill can be expressed in such a structured form and that a suitable verifier is available, which limits its general applicability.
On the other hand, without explicit behavioral evidence, earlier revisions are difficult to reassess when a problem recurs.
Several recent approaches retain historical records across evolution loops, but these records capture only scores or task-level status rather than linking targeted behaviors to edited regions~\cite{skillopt,skilladam}.
Such coarse records are insufficient to distinguish an incorrect edit location from an insufficient correction at the right location.
As a result, later loops may repeatedly modify the same content without resolving the deficiency, or fail to notice a regression introduced by an earlier revision.

\emph{Second}, it is challenging to distinguish deficiencies of the skills from incidental behaviors of the agent.
A deficiency is a property of the skills themselves and is shared by every execution under them, whereas agent behavior is stochastic and tasks differ in their conditions.
However, existing approaches typically attribute problems from limited evidence.
For example, SkillAdaptor attributes a failed trajectory to its earliest actionable step~\cite{skilladaptor}, and other approaches prompt an LLM to reflect on a minibatch of rollouts~\cite{skillopt,skilladam}.
In both cases, the LLM has to judge which observed problems are real, without a measure of how consistently a behavior is associated with failures across independent runs.
Consequently, revisions can fit individual cases without addressing deficiencies shared by other tasks of the same class, which limits the improvement on unseen tasks.

\emph{Third}, existing approaches struggle to address deficiencies that span multiple skills.
They generally revise only one skill or one document at a time.
For example, SkillAdaptor revises only the highest-weighted skill identified by its attribution mechanism~\cite{skilladaptor}, and SkillOpt and SkillAdam edit a single skill document~\cite{skillopt,skilladam}.
In practice, however, a task often involves several skills whose contents jointly guide the agent's behavior.
Consequently, updating one artifact in isolation can leave erroneous or conflicting content elsewhere, allowing the original problem to persist or recur.

\noindent\textbf{Our Work.}
In this work, we propose to evolve skills via \textit{trajectory-guided fault localization}, which identifies the skill contents that require revision.
The key idea is that before generating any revision, we explicitly determine the edit-sites, i.e., the skill contents that should be revised, and ground this decision in behavioral evidence from execution trajectories.
This idea draws on program repair, where evidence from passing and failing tests guides the localization of suspicious code before candidate patches are generated~\cite{sbfl-survey,genprog}.
Analogously, we treat each agent execution on a task as a test execution and the actions in its trajectory as the entities covered by that execution.
We localize through actions rather than treating skill contents directly as covered entities, because natural-language guidance lacks the explicit execution semantics of code.
Specifically, an agent loads a skill document as a whole, and whether and how it follows each instruction is left to its own interpretation, thus there is no observable notion of which contents an execution exercises.
In contrast, actions are directly observable in trajectories, differ across executions, and are shaped by the skill content.
Therefore, a deficiency in a skill manifests as suspicious actions, and the trajectory records the guidance the agent consulted when taking them, allowing these actions to be traced back to the responsible skill content.
By recording these actions, the edit sites derived from them, and the evidence behind both, the localization decisions become inspectable and can be reassessed when later executions reveal a persistent problem.
An action also provides a common anchor for locating regions across multiple skills that jointly guide it, thus enabling coordinated revisions to those contents.

We realize this idea in \tech.
\tech first abstracts raw trajectories into sequences of state transitions, making actions and their observed effects explicit units for inspection and comparison across executions.
Using these abstractions, \tech localizes edit sites in two stages before generating each revision.
The first stage identifies suspicious actions, i.e., those more likely to reflect deficiencies in the skills than incidental agent behaviors.
Our key insight is that reliable skill evolution requires both stability, by building on effective revisions across evolution loops to avoid recurring failures, and generality, by targeting deficiencies that persist across executions and tasks.
Accordingly, \tech progressively aggregates behavioral evidence across executions, evolution loops, and tasks.
Specifically, it first compares repeated executions of the same task and contrasts how each action occurs in failing and passing executions, 
where an action that introduces an error that is later recovered from in a successful execution also counts as failure evidence.
It then examines how the evidence changes across evolution loops to reveal whether earlier revisions took effect, and finally aggregates the evidence across tasks to prioritize problems shared by multiple tasks.
The second stage maps each suspicious action to edit sites in the skills.
Our observation here is that although a trajectory does not directly reveal which instructions the agent followed, it records the skills and files consulted, narrowing the search for guidance that may have shaped an action.
Based on this, an LLM-based agent uses the action's execution context to progressively narrow the search from skills to files to specific skill content, and outputs the edit sites along with a rationale linking the skill content to the action.
Finally, the edit sites, their localization rationale, and the supporting behavioral and historical evidence guide the generation of revision proposals, which are evaluated and consolidated into a patch that evolves the skill set.

\noindent\textbf{Results.}
We evaluate \tech on a repository-level software engineering skill benchmark SWE-Skills-Bench and a complex kernel-generation skill system CannBot.
Compared with the original skills, \tech achieves relative improvements of 15.84\% in pass@3 and 30.03\% in pass$^3$ on SWE-Skills-Bench, and 16.67\% and 100\% on CannBot, respectively.
\tech also achieves better overall performance than the four compared skill-evolution methods.
Moreover, in a deployment with our collaboration partner, \tech produced six skill-revision pull requests that have been accepted for future AI operator development, further showing its practical value.

\noindent\textbf{Summary.}
The main contributions of this paper are as follows:

\noindent\(\bullet\)~\textbf{Approach.}
We propose \tech, a skill-evolution approach based on trajectory-guided edit-site localization that links execution evidence to skill contents before generating revisions.

\noindent\(\bullet\)~\textbf{Evaluation.}
We conduct an empirical study on two software engineering skill benchmarks, demonstrating the effectiveness of \tech. 
In particular, \tech has been deployed in an AI operator development team, further confirming its practicality.

\section{Preliminaries}\label{sec:preliminary}

\subsection{Code Agent with Skill}\label{sec:background}

\begin{wrapfigure}{R}{0.4\linewidth}
    \centering
    \includegraphics[width=\linewidth]{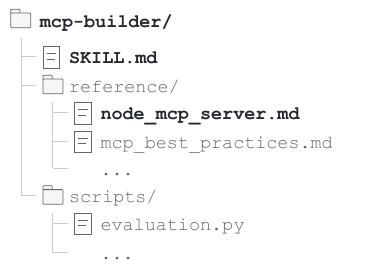}
    \caption{Structure of the \texttt{mcp-builder} skill.}
    \label{fig:mcp-skill-structure}
\end{wrapfigure}

In this work, we focus on code agents~\cite{openhands,swe-agent} that use skills for software engineering tasks and refer to the collection of skills available to an agent as its \emph{skill set}.
A skill set provides procedural knowledge and project-specific context for a class of related tasks without modifying the underlying model parameters~\cite{skillsbench,agent-skills-survey}.
Following the Agent Skills specification~\cite{agent-skills-specification}, each skill is a file-based package with a \texttt{SKILL.md} document as its entry point, as illustrated in Figure~\ref{fig:mcp-skill-structure}.
The \texttt{SKILL.md}'s front matter specifies the skill's name and description, while its body provides natural-language instructions.
The package may also contain supporting resources, such as reference documents, templates, and scripts, referenced by path.
Throughout this paper, \emph{skill} denotes the entire package, and \emph{artifact} denotes an individual file within it.

\noindent\textbf{Workflow of Code Agents.}
In general, a code agent accesses its skill set through progressive disclosure.
At startup, only the name and description of each installed skill are included in the system prompt.
When a task \(t\) arrives, the agent uses these descriptions to select relevant skills and loads their \texttt{SKILL.md} documents into context, and it reads supporting artifacts as needed during execution.
At each step, the agent selects an action (such as reading a file, editing code, or running tests) based on the loaded instructions and accumulated context, executes it, and receives the resulting observation.
This cycle continues until the task is completed or a resource limit is reached, yielding the execution trajectory
{
\begin{equation}
\tau=\langle \xi_1,\xi_2,\ldots,\xi_T\rangle,
\qquad
\xi_k=\langle a_k,o_k\rangle,
\label{eq:execution-trajectory}
\end{equation}
}%
where \(a_k\) is an action (e.g., a tool invocation) and \(o_k\) is the observation obtained after executing it.
A task-specific evaluation function $\Omega_t$ then maps the trajectory to an outcome $y=\Omega_t(\tau)\in\mathcal{Y}_t$, such as a pass/fail result or a structured record of functional correctness or performance.

As shown in the workflow, the skills affect the outcome through the actions that the agent selects under their guidance.
As a result, a deficiency in the skill cannot be observed from its content alone since the agent implicitly decides whether and how to follow each instruction.
Likewise, the final outcome summarizes an execution without necessarily identifying which actions caused a failure, and successful executions may include erroneous actions from which the agent later recovered.
Therefore, a deficiency only manifests as problematic actions in trajectories, which tend to recur across the related tasks that the skill set serves.
This makes actions the natural intermediate between outcomes and skill contents, and motivates evolving skills by first localizing suspicious actions and then tracing them back to the skill contents that guided them.

\subsection{Skill Evolution Problem}

\noindent\textbf{Quality of a Skill Set.}
In this paper, we study how to iteratively evolve a skill set to improve the effectiveness of a code agent.
Following this goal, we assess a skill set extrinsically, by the behavior it induces in the code agent that reads it.
Specifically, let \(\mathcal{A}\) denote a code agent equipped with a skill set \(\mathcal{S}\), and let \(D\) denote a set of tasks of the class that \(\mathcal{S}\) is intended to support.
An execution of $\mathcal{A}$ on a task $t\in D$ under $\mathcal{S}$ produces a record \((\tau,y)\), as stated in Equation~\ref{eq:execution-trajectory}, where \(\tau\) is the execution trajectory and \(y=\Omega_t(\tau)\) is the outcome of \(\tau\).
Let \(\mathcal{Y}^{+}_t\subseteq\mathcal{Y}_t\) denote the outcomes in which task \(t\) is solved.
The quality of \(\mathcal{S}\) over \(D\) is then defined as the expected proportion of tasks solved,
{
\begin{equation}
J_{D}(\mathcal{S})=\frac{1}{|D|}\sum_{t\in D}\;
\mathbb{E}_{\tau\sim\mathcal{A}(t,\mathcal{S})}
\bigl[\mathbf{1}\{\Omega_t(\tau)\in\mathcal{Y}^{+}_t\}\bigr].
\label{eq:skill-quality}
\end{equation}
}%

\noindent\textbf{Skill Evolution.}
Given a code agent \(\mathcal{A}\) whose model and harness are fixed, an initial skill set \(\mathcal{S}_0\), an evolution task set \(D_e\), and a loop budget \(L\).
In loop \(i\), executing the tasks in \(D_e\) under \(\mathcal{S}_i\) yields a set of execution records \(\mathcal{E}_i=\{(\tau,y)\}\), from which the next skill set is produced.
Skill evolution aims to construct this revision procedure so that the skill set obtained after \(L\) loops is as effective as possible on unseen tasks:
{
\begin{equation}
\max\;J_{D_v}(\mathcal{S}_L)
\qquad\text{s.t.}\qquad
\mathcal{S}_{i+1}=\mathrm{Evolve}\bigl(\mathcal{S}_i,\mathcal{E}_i\bigr),
\quad 0\leq i<L,
\label{eq:problem}
\end{equation}
}%
where each \(\mathcal{S}_{i+1}\) differs from \(\mathcal{S}_i\) only in the contents of the artifacts, and \(D_v\) is a held-out set of tasks of the same class as \(D_e\) that is never accessed during evolution, i.e., \(D_e\cap D_v=\emptyset\).

Equations~\ref{eq:skill-quality} and~\ref{eq:problem} motivate three design considerations for the function \(\mathrm{Evolve}\).
First, \(\mathrm{Evolve}\) should distinguish persistent skill deficiencies from incidental agent behavior.
Equation~\ref{eq:skill-quality} defines \(J_D\) as an expectation over executions averaged across related tasks, thus a failure in one execution or task need not indicate a persistent deficiency.
Moreover, \(\mathrm{Evolve}\) observes executions only on \(D_e\), whereas its objective is evaluated on \(D_v\).
Thus, revisions that fit incidental behavior on \(D_e\) can improve \(J_{D_e}\) without improving \(J_{D_v}\), motivating the identification of deficiencies that persist across executions and tasks.
Second, \(\mathrm{Evolve}\) should retain and reuse historical evidence to preserve gains across evolution loops.
Since \(\mathcal{S}_L\), rather than the most effective intermediate skill set, is returned, a revision that disrupts earlier effective guidance leaves a regression carried to \(\mathcal{S}_L\).
Linking each revision to its motivating behavioral evidence and subsequent executions supports reassessing earlier edits and detecting these regressions.
Third, \(\mathrm{Evolve}\) should support coordinated revisions across multiple skills.
A task may involve several skills whose contents jointly guide the agent's actions.
Therefore, producing an improved \(\mathcal{S}_{i+1}\) may require revising multiple skills, since an isolated edit can leave erroneous or conflicting guidance elsewhere.
Together, these considerations motivate revising the skill contents responsible for persistent deficiencies, potentially across multiple skills, using behavioral evidence accumulated across executions, evolution loops, and tasks.

\section{\tech in Action}
\label{sec:in-action}

\begin{figure}[!tb]
    \centering
    \captionsetup[subfigure]{font=scriptsize,labelfont={sf,scriptsize},textfont={sf,scriptsize},skip=3pt,justification=centering}
    \begin{subfigure}[t]{\linewidth}
        \centering
        \includegraphics[width=\linewidth]{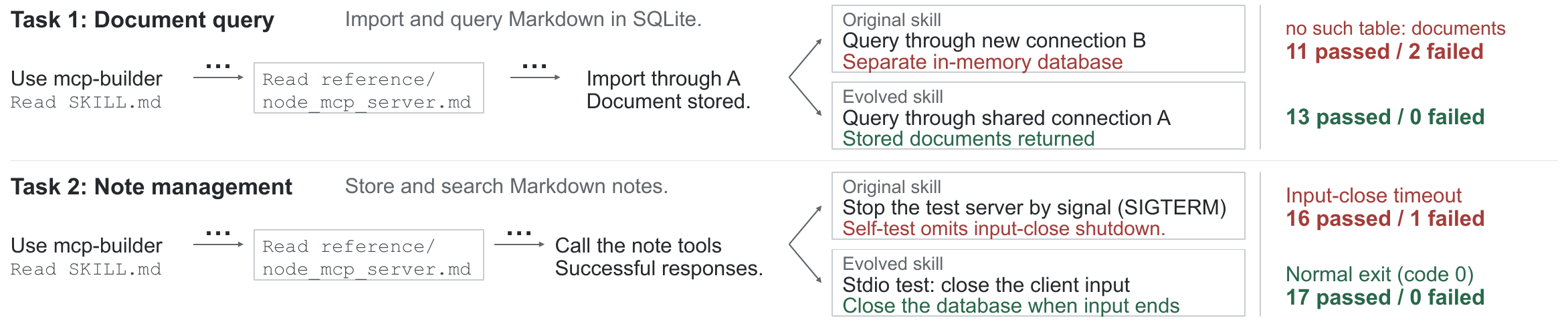}
        \caption{Task executions before and after skill evolution.}
        \label{fig:case-mcp-errors}
    \end{subfigure}
    \begin{subfigure}[t]{\linewidth}
        \centering
        \includegraphics[width=\linewidth]{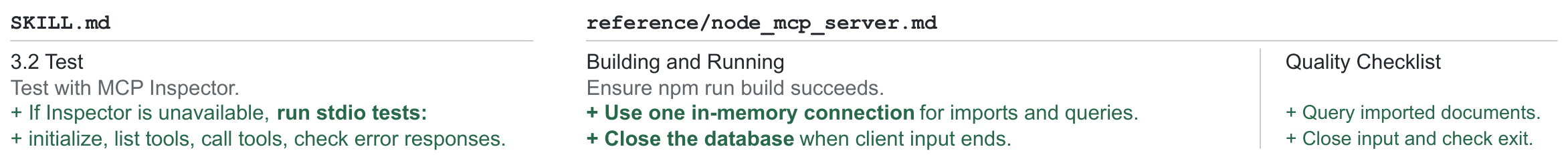}
        \caption{Coordinated skill revision.}
        \label{fig:case-mcp-revision}
    \end{subfigure}
    \caption{An example of \tech evolving the \texttt{mcp-builder} skill.}
    \label{fig:case-mcp}
\end{figure}

We use the \texttt{mcp-builder} skill to show how \tech turns failures on different tasks into one coordinated revision across skill artifacts.
The skill guides agents in building Model Context Protocol (MCP) servers that expose application functions as tools; as Figure~\ref{fig:mcp-skill-structure} shows, it consists of a workflow in \texttt{SKILL.md}, implementation references such as \path{reference/node_mcp_server.md}, and testing scripts.
For illustration, we use document query (Task~1) and note management (Task~2) for evolution, and hold out Task~3, which indexes Markdown documents for search and retrieval.

\noindent\textbf{Failures under the original skill.}
Figure~\ref{fig:case-mcp}(\subref{fig:case-mcp-errors}) compares key execution fragments and verification results before and after evolution.
In Task~1, the agent imports a document through connection~A but queries through a second in-memory connection~B.
Since an in-memory SQLite database is private to its connection, B cannot see the imported data, and verification fails with \texttt{no such table: documents} (11/13 checks pass).
In Task~2, the agent's self-test succeeds and ends by stopping the server with \texttt{SIGTERM}, never checking shutdown on input close.
However, verification does: the server keeps running after the client's input closes, and the check times out (16/17 checks pass).

\noindent\textbf{Localization.}
In both tasks, the problematic action (opening connection~B; stopping the server by \texttt{SIGTERM}) follows a read of \path{reference/node_mcp_server.md}, thus \tech traces both to this artifact, whose \emph{Building and Running} and \emph{Quality Checklist} sections have no rule for sharing in-memory connections or cleaning up on input close.
Additionally, the testing gap in Task~2 further leads \tech to the \emph{3.2 Test} section of \texttt{SKILL.md}, which recommends MCP Inspector but gives no fallback.

\noindent\textbf{Coordinated revision.}
As Figure~\ref{fig:case-mcp}(\subref{fig:case-mcp-revision}) shows, \tech revises these three sections together: \emph{Building and Running} gains two runtime rules (use one in-memory connection; close the database when input ends), \emph{Quality Checklist} gains the matching checks, and \emph{3.2 Test} gains a stdio testing procedure (initialize, list tools, call tools, check error responses) that keeps the checks runnable without Inspector.
Revising either artifact alone would leave the guidance incomplete: the checks would lack a procedure to run them, or the procedure would lack rules and checks to apply.

\noindent\textbf{Outcome.}
With the evolved skill, the agent queries through the shared connection~A in Task~1 and closes the database when input ends in Task~2, passing all checks.
To check that the revision repairs the guidance rather than fitting these two tasks, we evaluate on the held-out Task~3.
Task~3 exhibits the same query and shutdown failures under the original skill; with the evolved skill, retrieval succeeds and the server terminates normally.

\section{Approach}
\begin{figure*}[!tb]
    \centering
    \includegraphics[width=\linewidth]{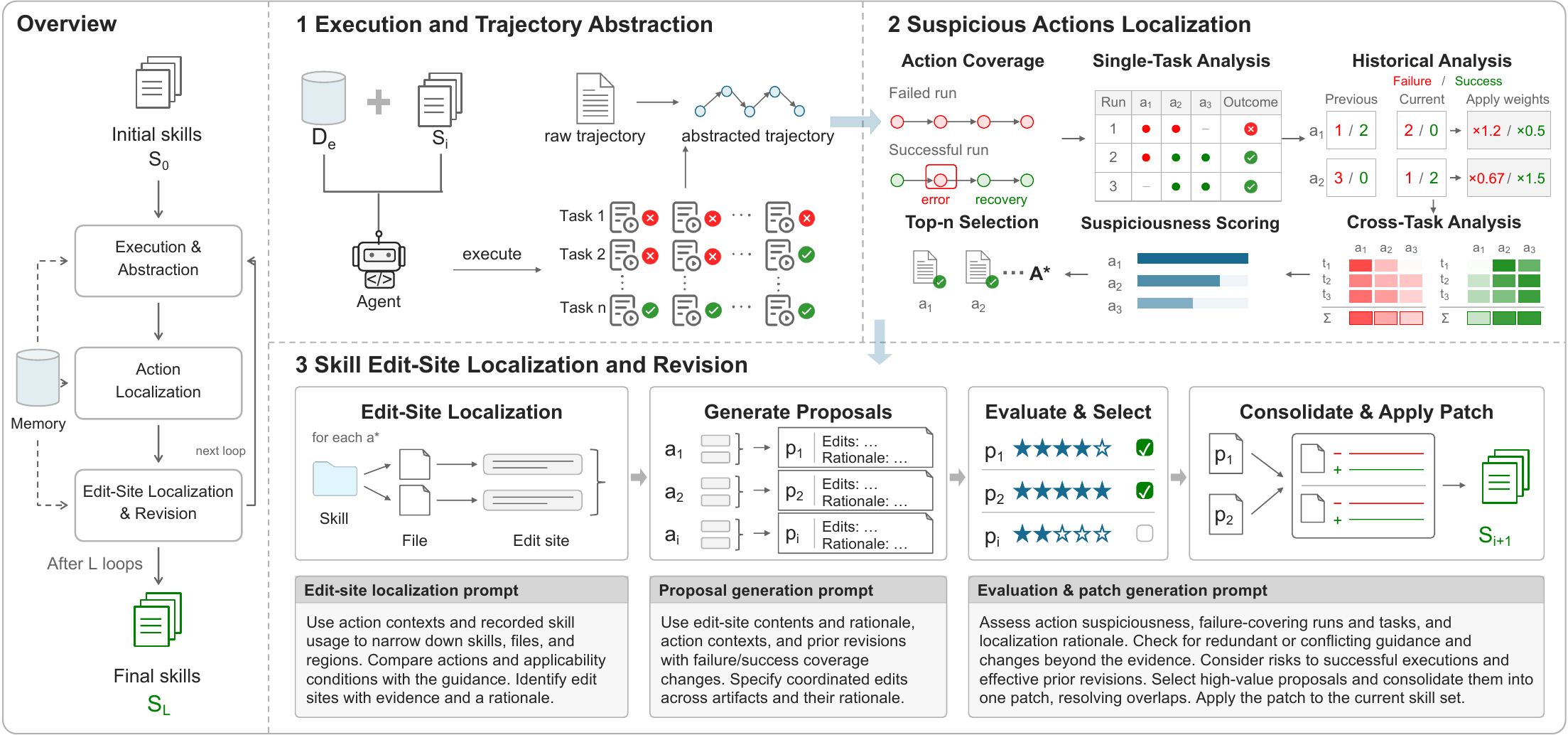}
    \caption{Overview of \tech.}
    \label{fig:overview}
\end{figure*}

Figure~\ref{fig:overview} illustrates the workflow of \tech.
Given an initial skill set \(\mathcal{S}_0\) and a task set \(D_e\), \tech evolves the skills over \(L\) loops, each comprising three stages.
First, \textit{Execution and Trajectory Abstraction} repeatedly executes the tasks with the current skill set and abstracts the resulting trajectories into a state-transition representation, enabling comparison across executions.
Using these abstractions, \textit{Suspicious Actions Localization} first compares the failure and success coverage of actions across repeated executions of each task.
Then, it adjusts this evidence according to changes since the preceding loop, and finally aggregates it across tasks to score and select suspicious actions.
These suspicious actions guide \textit{Edit-Site Localization and Revision}, which locates the relevant skills, artifacts, and contents, generates and evaluates revision proposals, and applies a patch that consolidates the selected proposals to evolve the current skill set.
In addition, an evolution memory preserves execution evidence, analyses, and revision history to support these analyses.
After \(L\) loops, \tech returns the resulting skill set \(\mathcal{S}_L\).

\subsection{Execution and Trajectory Abstraction}
\label{sec:executor}

This stage collects execution evidence under the current skill set to support subsequent edit-site localization and revision generation.
\tech first executes each evolution task multiple times and retains all runs, regardless of their outcomes.
It then abstracts each raw trajectory into a uniform state-transition representation that captures actions and their observed effects, enabling analysis within individual runs and comparison across executions.
We detail these two steps below.

\noindent\textbf{Task Execution}\label{sec:task-execution}
To obtain a set of execution trajectories, we independently run the code agent multiple times on each evolution task under the current skill set.
Repeated execution is important to \tech because agent behavior is stochastic: a single run cannot separate a deficiency induced by the skills from an incidental sampling artifact~\cite{process-centric-agents, code-agent-behaviour,failure-process}.
Formally,  in evolution loop \(i\), the fixed code agent \(\mathcal{A}\) executes each task \(t \in D_e\) independently \(m\) times under the current skill set \(\mathcal{S}_i\), producing
\(
\mathcal{T}_i(t)
=
\{
(\tau_{i,t}^{(j)}, y_{i,t}^{(j)})
\}_{j=1}^{m},
\)
where \(\tau_{i,t}^{(j)}\) and \(y_{i,t}^{(j)}\) denote the trajectory and final outcome of the \(j\)-th run, respectively.
Moreover, \tech retains all trajectories, regardless of their outcomes, because successful and failed runs provide complementary evidence.
A successful run may involve repeated unsuccessful attempts before recovery, revealing potential weaknesses in skill guidance that the final outcome alone does not expose.
Conversely, a failed run may contain effective intermediate actions that remain useful despite the overall failure.

\noindent\textbf{Trajectory Abstraction}
Given the collected execution trajectories \(\mathcal{T}_i(t)\), a straightforward approach is to provide them directly to an LLM for failure attribution or revision generation, as done in prior work~\cite{skilladam, skilladaptor, evoskill}.
However, raw trajectories present two challenges for analysis.
First, record boundaries do not necessarily correspond to meaningful behaviors.
A single agent step may contain several tool calls whose responses must be examined together, while individual records may provide little information about the behavior being analyzed.
The analysis therefore requires units that capture meaningful operations and their effects.
Second, raw trajectories are lengthy and verbose, and similar behaviors may differ in wording, ordering, and number of steps across runs, making them difficult to compare.

To combat these two challenges,
we abstract each raw trajectory into a uniform state-transition representation, drawing on prior work on action--state mapping and execution-log
normalization~\cite{famas,codetracer}.
More specifically, for each trajectory \(\tau_{i,t}^{(j)} \in \mathcal{T}_i(t)\), we follow prior work~\cite{famas,codetracer} to use an LLM equipped with a hierarchical clustering algorithm to semantically parse and normalize the trajectory \(\tau_{i,t}^{(j)}\) into an ordered sequence of transitions \((\mu_{k-1}, a_k, \mu_k)\).
Here, \(a_k\) describes the action and affected object, while \(\mu_{k-1}\) and \(\mu_k\) describe the observed states before and after the action, respectively.
This yields the abstracted trajectory set \(\widehat{\mathcal{T}}_i(t) = \{(\hat{\tau}_{i,t}^{(j)}, y_{i,t}^{(j)})\}_{j=1}^{m}\),
where \(\hat{\tau}_{i,t}^{(j)} = \langle \mu_0, a_1, \mu_1, \ldots, a_K, \mu_K \rangle\) and \(K\) denotes the number of abstracted operations.
Finally, by aggregating all trajectories from different tasks, we obtain the execution trajectory set \(\widehat{\mathcal{T}}_i=\bigcup_{t\in{D_e}}\widehat{\mathcal{T}}_i(t)\) for the current skill set \(\mathcal{S}_i\).

\subsection{Suspicious Actions Localization}
This stage aims to identify the suspicious actions from the execution trajectory set \(\widehat{\mathcal{T}}_i\) that are more likely to reflect deficiencies in the current skill set \(\mathcal{S}_i\).
The key premise is that a deficiency is a property of the skills themselves, which every execution under the skill set shares, thus the actions it induces tend to recur across independent runs and tasks.
However, routine actions, such as reading files or running tests, also recur during normal execution.
Therefore, recurrence alone is insufficient; we must compare how actions occur in favorable and unfavorable execution contexts.
This follows the intuition of spectrum-based fault localization (SBFL), which ranks program entities by comparing their coverage in failing and passing tests, namely entities consistently covered by failing tests but seldom by passing tests warrant closer inspection~\cite{sbfl-survey}.
We adapt this principle by treating actions as the entities to be localized and individual runs as executions.
Note that FAMAS adopts a related approach to failure attribution in multi-agent systems by comparing repeated executions of the same task~\cite{famas}.
However, skill evolution additionally requires identifying problems shared across tasks and tracking how behavioral evidence changes as skills are revised.
Therefore, we adjust each task's spectrum counts according to their changes since the preceding loop and aggregate the adjusted evidence across tasks.

For a task \(t\in{D}_e\) in loop \(i\), let \(\widehat{\mathcal T}_i(t)\) denote the set of its \(m\) runs, each comprising an abstracted trajectory \(\hat\tau\) and its outcome \(y\).
We say that a run covers an action \(a\) if its trajectory contains at least one occurrence of \(a\), i.e., \(a_k=a\) for some step \(k\).
We then count the runs that provide failure and success coverage of \(a\) according to the following rules.
If \(y\) indicates a failure, every action occurring in the run directly receives failure coverage, as a statement executed by a failing test does in SBFL.
If \(y\) indicates a success, we further inspect whether an action introduced an error that the agent later recovered from during the execution.
This inspection is necessary because LLM-based agents typically repair their own errors, so a successful run may still contain errors that an action introduced and the agent later recovered from.
However, such recovery varies across runs: the same error that is repaired in one run may remain unresolved in another, causing the run to fail.
Counting every action of a successful run as success coverage would therefore exonerate the very actions that cause failures elsewhere.
Moreover, even when the recovery succeeds, an action that introduces an error still reveals a deficiency, since the skill should ideally guide the agent to avoid both the error and the cost of recovering from it.
We thus inspect each occurrence of an action at step \(k\) in a successful run.
Specifically, we adopt an LLM to examine the window of \(h\) steps before and after \(k\), i.e., the states and actions from \(\mu_{k-h-1}\) to \(\mu_{k+h}\), and thus to decide whether the action introduced an error state (an unresolved error or a failed validation) rather than merely revealing an existing one.
If so, the occurrence contributes failure coverage; otherwise, it contributes success coverage.
Note that this LLM-based judgment is deliberately narrow: it asks only whether the states around an action show an error that the action introduced, not which action caused a failed outcome, a question that requires reasoning over the whole trajectory and remains difficult even for strong LLMs~\cite{who-when}.
Finally, each run contributes at most once to each coverage count of \(a\), so that the resulting spectrum measures how many independent executions exhibit the action rather than how often it repeats within a single run.

\noindent\textbf{Single-Task Analysis.}
We first measure the failure and success coverage of each action across the \(m\) runs of task \(t\), reducing sensitivity to stochastic variation in agent execution.
Within this comparison, a suspicious action should meet two conditions.
First, an action should occur across runs providing failure evidence, giving stronger support than an isolated occurrence.
Second, its failure coverage should be considered relative to its success coverage, helping distinguish problematic actions from routine behavior.
Using only the number of failure-covering runs favors routine actions, such as reading files or running tests, which appear in nearly every run regardless of the outcome.
Conversely, using only the share of failure coverage among all coverage of an action can overemphasize an action observed in a single failed run.
This is analogous to the Ochiai form~\cite{Ochiai} widely used in SBFL.
We therefore follow the Ochiai form to compute the spectrum matrix.
Formally, let \(\chi^-(\hat{\tau},y;a)\) and \(\chi^+(\hat{\tau},y;a)\) indicate whether a run provides failure and success coverage of \(a\), respectively.
Let \(\eta^-(\hat{\tau},y)\) indicate whether the run provides failure coverage of at least one action.
The task-level counts are:
\begin{equation}
\label{eq:task-behavior-coverage}
N_{i,t}^{\sigma}(a)
=
\sum_{(\hat{\tau},y)\in\widehat{\mathcal{T}}_i(t)}
\chi^\sigma(\hat{\tau},y;a),
\qquad
F_{i,t}
=
\sum_{(\hat{\tau},y)\in\widehat{\mathcal{T}}_i(t)}
\eta^-(\hat{\tau},y),
\quad \sigma\in\{-,+\}.
\end{equation}
Here, \(N^-_{i,t}(a)\) and \(N^+_{i,t}(a)\) count the runs providing failure and success coverage of \(a\), respectively, while \(F_{i,t}\) counts all runs providing failure coverage of at least one action.

\noindent\textbf{Historical Analysis.}
The counts above describe how an action occurs in the current loop, but not how this evidence has evolved.
The same counts can arise from a deficiency that the preceding revision has partly resolved, one that it failed to fix, or a regression that it introduced, yet these cases call for different priorities.
To address this, we compare each task's current coverage counts with those of the preceding loop and adjust them accordingly.
The key insight is that, between two adjacent loops, only the skill set differs, owing to the revision made in the preceding loop.
The change in coverage counts can thus be read as an effect of that revision.
Specifically, we adjust failure and success coverages according to the direction of their change.
For failure coverage, a decrease suggests that the revision has taken effect, so we downweight the current failure evidence to avoid repeatedly revisiting problems that may already be improving.
An increase suggests that the revision was ineffective or misdirected, thus we upweight the evidence to highlight actions increasingly associated with failure.
This includes newly observed actions, whose failure coverage rises from zero and may indicate a regression introduced by the revision.
Success coverage serves as contrasting evidence in the suspiciousness calculation, and its change carries the reverse implication.
An action whose success coverage grows may be an effective practice that the revision has promoted, thus we upweight its success evidence to shield it from suspicion, whereas a decrease weakens this contrasting evidence.
When coverage remains unchanged or no comparable history is available, the adjustment is neutral, so a persistent deficiency retains its full weight while improving ones are downweighted.
Together, these adjustments account for both the current evidence and its direction of change across evolution loops.

Formally, for each action \(a\), we retrieve the preceding loop's records of the same task from the evolution memory and compare the coverage counts:
\begin{equation}
\label{eq:historical-coverage-ratio}
r_{i,t}^{\sigma}(a)
=
\frac{N_{i,t}^{\sigma}(a)+1}
     {N_{i-1,t}^{\sigma}(a)+1},
\qquad \sigma\in\{-,+\}.
\end{equation}
Adding \(1\) to each count accommodates zero coverage and softens the ratio when the counts are small.
Based on \(r_{i,t}^{\sigma}(a)\), we then borrow the bounded transformation used in Yule's \(Q\)~\cite{warrens2008} and shift it to obtain a multiplicative weight with a neutral value of \(1\):
\begin{equation}
\label{eq:historical-coverage-weight}
w_{i,t}^{\sigma}(a)
=
1+
\frac{r_{i,t}^{\sigma}(a)-1}
     {r_{i,t}^{\sigma}(a)+1}
=
\frac{2\bigl(N_{i,t}^{\sigma}(a)+1\bigr)}
     {N_{i,t}^{\sigma}(a)+N_{i-1,t}^{\sigma}(a)+2}.
\qquad \sigma\in\{-,+\}.
\end{equation}
Finally, we separately adjust failure and success coverage counts obtained from single-task analysis:
\begin{equation}
\label{eq:history-adjusted-coverage}
\widetilde N_{i,t}^{-}(a)
=
w_{i,t}^{-}(a)N_{i,t}^{-}(a),
\qquad
\widetilde N_{i,t}^{+}(a)
=
w_{i,t}^{+}(a)N_{i,t}^{+}(a).
\end{equation}
Note that such adjusted coverage counts are used only for identifying suspicious actions in the current loop and are not accumulated across loops.

\noindent\textbf{Cross-Task Analysis.}
As discussed in \S\ref{sec:preliminary}, a skill set typically supports a class of related tasks, indicating that their revisions should generalize to unseen tasks of the same type.
However, the adjusted coverage counts \(\widetilde N_{i,t}^{-}(a)\) and \(\widetilde N_{i,t}^{+}(a)\) are specific to a single task \(t\), and the actions they flag are likely bound to that task's conditions, thus providing limited signals for skill evolution.
Symmetrically, an action that stably appears in the successful runs of other tasks is harmless in itself, so its failures in this task more likely stem from the task's context.
Therefore, we aggregate the coverage counts after historical adjustment across tasks:
\begin{equation}
\label{eq:cross-task-behavior-evidence}
\overline N_i^{-}(a)
=
\sum_{t\in D_e}\widetilde N_{i,t}^{-}(a),
\qquad
\overline N_i^{+}(a)
=
\sum_{t\in D_e}\widetilde N_{i,t}^{+}(a).
\end{equation}
It is important to note that we keep the historical adjustment to each task before aggregation because aggregating first would mask the different trends across tasks, e.g., a decrease in failure coverage on one task may offset an increase on another, leaving the total unchanged.
In contrast, adjusting first instead lets each task's change shape its own contribution to the combined evidence.

\noindent\textbf{Suspicious Action Identification.}
Now we state how to identify suspicious actions based on the coverage counts. 
We keep every action with failure coverage in at least one task, i.e., \(\sum_{t\in D_e}N^-_{i,t}(a)>0\), as a candidate.
For each candidate, we measure its suspicious score as follows:
\begin{equation}
\label{eq:behavior-suspiciousness}
s_i(a)
=
\sqrt{
\underbrace{\frac{\overline N_i^{-}(a)}{F_i}}_{\text{coverage ratio}}
\cdot
\underbrace{\frac{\overline N_i^{-}(a)}{\overline N_i^{-}(a)+\overline N_i^{+}(a)}}_{\text{specificity ratio}}
}
=
\frac{\overline N_i^{-}(a)}
{\sqrt{F_i\left(\overline N_i^{-}(a)+\overline N_i^{+}(a)\right)}},
\qquad
F_i=\sum_{t\in D_e}F_{i,t}.
\end{equation}
In Equation~\ref{eq:behavior-suspiciousness}, the coverage ratio measures how widespread \(a\) is among the failure-covering runs across tasks, while the specificity ratio is the share of failure coverage in the total coverage of \(a\), and measures how specific \(a\) is to failures.
For example, an action that recurs in the failures of many tasks accumulates a large \(\overline N_i^{-}(a)\), whereas one confined to a single task covers only a fraction of the failure-covering runs and receives a small coverage ratio.
An action that also appears in the successful runs of other tasks accumulates a large \(\overline N_i^{+}(a)\) and receives a low specificity ratio.

Finally, we rank all candidates by their suspicious scores and select the top-\(n\) actions as the suspicious actions, denoted as \(A^\star\) for guiding the subsequent skill evolution. 

\subsection{Skill Edit-Site Localization and Revision}
\label{sec:proposal}

Based on the identified suspicious actions \(A^\star\), this stage first localizes the skill contents to be revised (i.e., edit sites) by linking each suspicious action to the skill content that guides it, and then generates revisions for these edit sites to evolve the skill set \(\mathcal{S}_i\).

\noindent\textbf{Edit-Site Localization.}
A skill set generally consists of natural-language instructions, together with reference documents, templates, and scripts spread over multiple skills and artifacts, which makes it hard to link an action to the contents that guide it by static rules.
We address this challenge with an LLM-based agent that reads the skill set and maps each of the suspicious actions \(a^\star\in{A^\star}\) to the edit sites in the current skill set.

For each \(a^\star\), the agent walks through the skill set \(\mathcal{S}_i\) recursively, narrowing from skills to artifacts to contents, rather than processing it all at once, since the latter would bring irrelevant contents (e.g., skills unrelated to \(a^\star\)) into its context and add noise to the localization.
Specifically, it first examines the occurrences of \(a^\star\) and their execution contexts in the trajectories retained during aggregation, alongside the skill descriptions and the recorded skill usage, to identify the skills that guide these actions.
Then, it inspects the \texttt{SKILL.md} files of these skills and the artifacts they reference to find the instructions or script logic relevant to the actions.
Next, it compares the prescribed actions and applicability conditions in these contents with the actions, revealing where the guidance prescribes unsuitable actions, omits necessary conditions, or conflicts with related contents, and accordingly these places are the edit sites.
Finally, the agent outputs an edit-site for \(a^\star\), together with the supporting execution evidence and a rationale that links the skill contents to the action.
In particular, an edit site specifies the set of skill contents to be revised, represented by their line ranges together with the corresponding skills and artifacts. 
Note that because a suspicious action may be associated with multiple skills or artifacts, a single edit site may span multiple contents across different skills or artifacts.

\noindent\textbf{Revision Generation and Application}
Based on the identified edit-sites, \tech next develops revision proposals and implements them as patches to evolve the skills.
Specifically, we first prompt an LLM to generate a revision proposal for each edit site, specifying the intended revisions and their rationale.
The proposal generation draws on three complementary sources of information.
First, the edit site with its current contents and the localization rationale, which reveal what the existing skill contents prescribe and why they fail, and thus guide how to revise the site without introducing redundancy or conflicts.
Second, the action \(a^\star\) associated with the edit site and its execution contexts in the trajectories, which specify the problem that the revision should address.
Third, the revision history, i.e., the earlier revisions to the relevant skills together with the changes in the failure and success coverage of \(a^\star\) between adjacent loops (measured by Eq.~\ref{eq:historical-coverage-ratio}), which show whether the earlier edits took effect, so that the new revision can retain, refine, or revert them instead of repeating ineffective changes.
In addition, we explicitly specify the structure of each revision proposal, including the coordinated changes required across the involved skills or artifacts to maintain consistency.

The proposals of different edit sites can overlap, and not all of them are worth applying (e.g., some duplicate existing skill contents).
Therefore, we evaluate the proposals and select a subset for developing the final patch, instead of applying them directly.
Specifically, we design a chain-of-thought prompt that guides the LLM through three steps.
First, the LLM evaluates the proposals in three aspects:
(1) the evidence of the proposal's action \(a^\star\), i.e., its suspiciousness score \(s_i(a^\star)\) and the runs and tasks in which it provides failure coverage, together with the localization rationale, which indicate how strongly the problem is established, how many failures the revision can address, and how certain the localization is;
(2) the proposal and the current skill contents around its edit sites, which show whether the change duplicates or conflicts with the existing guidance and whether it extends beyond what the evidence supports; and
(3) the successful cases and the revision history, i.e., the successful runs that rely on the edited contents and the observed effects of earlier revisions, which show whether the change may disrupt guidance supporting successful executions or undo an effective earlier revision.
Based on the evaluation, the LLM then selects the high-value proposals.
Finally, it consolidates the selected proposals into the final patch, resolving any overlap among them.
Applying the patch evolves the current skill set \(\mathcal{S}_i\) into \(\mathcal{S}_{i+1}\), which is used in the next evolution loop or returned as the final skill set \(\mathcal{S}_L\) when evolution completes.

\section{Experiment}
\label{sec:experiment}
We evaluate \tech on the following research questions:

\noindent\(\bullet\) \textbf{RQ1:} How effective and efficient is \tech in evolving agent skill sets?

\noindent\(\bullet\) \textbf{RQ2:} How do skills evolved by \tech compare in effectiveness with the original skills and those evolved by existing methods?

\noindent\(\bullet\) \textbf{RQ3:} How do the configurations and key components of \tech affect its effectiveness?

\subsection{Experimental Setup}
\label{sec:experimental-setup}
\noindent\textbf{Benchmarks.}
We evaluate \tech on skill sets that vary in structure and complexity: independent skills from SWE-Skills-Bench~\cite{swe-skills-bench} and the coordinated multi-skill system of CannBot.

\noindent\(\bullet\) \emph{SWE-Skills-Bench.}
It comprises 49 independent skills for software engineering, each paired with ten different task batches grounded in a real-world GitHub repository pinned to a fixed commit.

\noindent\(\bullet\) \emph{CannBot.}
It comprises 12 coordinated skills for complex AscendC kernel generation in the CANN ecosystem~\cite{ascend-architecture}.
We evaluate CannBot on tasks drawn from NPUKernelBench~\cite{npubench}.

\noindent\textbf{Baselines.}
The baselines include the original skills and four existing skill-evolution methods.

\noindent\(\bullet\) \emph{Original Skills.}
We use the carefully curated public skills from SWE-Skills-Bench and the official CannBot skill system designed for AscendC kernel generation as the unevolved baselines.

\noindent\(\bullet\) \emph{Skill-Evolution Approaches.}
We select comparison methods based on three criteria:
{(1)} Their source code is publicly available, enabling reproduction and adaptation.
{(2)} They demonstrate strong performance in their original evaluation settings.
{(3)} Their skill-evolution workflows can be adapted to our benchmarks by integrating the corresponding task-execution and verification procedures.
Based on these criteria, we select four representative skill-evolution approaches.
\emph{SkillOpt}~\cite{skillopt} directly optimizes a single skill document through rollout-guided textual edits and validation gating.
\emph{SkillAdaptor}~\cite{skilladaptor} first localizes an actionable fault in a failed trajectory, then revises a responsible skill or generates a new one when skill coverage is missing.
\emph{EvoSkill}~\cite{evoskill} evolves complete structured skill folders through failure-driven mutations and Pareto-frontier selection based on validation performance.
\emph{CAP-Evolve}~\cite{capevolve} evolves agent prompts, skill packages, and tools through an iterative workflow of failure diagnosis, candidate generation, and validation-based selection.

\noindent\textbf{Metrics.}
We use three metrics to evaluate effectiveness.
A trial is considered successful only if the agent completes the task within the time limit and passes all tests or verification checks provided by the benchmark.
Following prior work on code generation and agent evaluation~\cite{passk,taubench}, we use \emph{pass@3} and \emph{pass$^3$} to measure the proportions of tasks solved in at least one of three independent trials and in all three trials, respectively.
In addition, we use \emph{trial-level accuracy} to measure the proportion of successful individual trials.

\noindent\textbf{Implementation.}
We implement \tech in Python 3.11 following its three-stage workflow.
We use Claude Code~\cite{claude-code} with DeepSeek-V4-Flash-0731~\cite{deepseek-v4} as the code agent for task execution using skills, and the same model throughout evolution.
To reduce wall-clock time, we parallelize task-level analyses across independent tasks.
Unless otherwise stated, each evolution task is independently executed \(m=3\) times per loop, with \(L=3\) loops for SWE-Skills-Bench and \(L=5\) for CannBot.
Task execution follows SWE-Skills-Bench's timeout settings, while CannBot uses a 9,000-second limit.
Each loop selects the top-10 suspicious actions as inputs to edit-site localization.

To assess generalization to tasks unseen during evolution, the tasks are divided into an evolution set \(D_e\), which provides execution evidence for skill evolution, and a held-out evaluation set \(D_v\), which is never accessed during evolution.
For each skill in SWE-Skills-Bench, we partition its ten tasks into two five-task folds, A and B, containing tasks with odd- and even-numbered original IDs, respectively.
We conduct two complementary experiments, A\(\rightarrow\)B and B\(\rightarrow\)A, by swapping the folds' roles as \(D_e\) and \(D_v\), with both experiments starting from the same original skill.
Because CannBot incurs substantially higher execution costs, we use a single fixed split with 15 tasks for evolution and 15 for held-out evaluation.
For each set, an LLM selects 10 Level-1 and 5 Level-2 tasks from NPUKernelBench, prioritizing task diversity.

We reproduce the compared skill-evolution methods using their official implementations, with the same settings as \tech:
(1) code agent and task split;
(2) execution results from the original skills for the first evolution loop;
and (3) numbers of executions per task and evolution loops.
We adapt their execution and verification interfaces while retaining their evolution workflows.

\noindent\textbf{Environment.}
The SWE-Skills-Bench experiments run on an Ubuntu 22.04.4 machine with an Intel Xeon Gold 6426Y CPU and 512\,GB RAM.
The CannBot experiments run on a machine with eight Ascend 910B3 NPUs and 1.5\,TB RAM, using CANN 9.1.0.

\subsection{RQ1: Overall Performance of \tech}
\label{sec:rq1}

\begin{table}[!tb]
    \centering
    \caption{Pass@3, trial-level accuracy, and Pass$^3$ (\%) of the original skills, skill-evolution baselines, and \tech on SWE-Skills-Bench and CannBot.}
    \label{tab:main-results}
    \fontsize{8}{9.6}\selectfont
    \setlength{\tabcolsep}{2.0pt}
    \renewcommand{\arraystretch}{1.10}
    \setlength{\arrayrulewidth}{0.4pt}
    \begin{tabular*}{\linewidth}{@{\extracolsep{\fill}}l*{9}{c}|*{3}{c}@{}}
        \toprule
        \multirow{2}{*}{Method}
        & \multicolumn{3}{c}{SWE: A$\rightarrow$B}
        & \multicolumn{3}{c}{SWE: B$\rightarrow$A}
        & \multicolumn{3}{c}{SWE: Overall}
        & \multicolumn{3}{|c}{CannBot} \\
        \cmidrule(lr){2-4}\cmidrule(lr){5-7}\cmidrule(lr){8-10}\cmidrule(l){11-13}
        & Pass@3 & Trial Acc. & Pass$^3$
        & Pass@3 & Trial Acc. & Pass$^3$
        & Pass@3 & Trial Acc. & Pass$^3$
        & Pass@3 & Trial Acc. & Pass$^3$ \\
        \midrule
        Base Skills
        & 84.49 & 77.69 & 71.02
        & 80.41 & 73.20 & 64.90
        & 82.45 & 75.44 & 67.96
        & 80.00 & 66.67 & 46.67 \\
        SkillOpt
        & 93.47 & 89.93 & 85.71
        & 84.49 & 79.86 & 75.10
        & 88.98 & 84.90 & 80.41
        & 80.00 & 71.11 & 53.33 \\
        EvoSkill
        & 87.76 & 83.27 & 78.37
        & 83.67 & 77.55 & 70.61
        & 85.71 & 80.41 & 74.49
        & 86.67 & 73.33 & 53.33 \\
        SkillAdaptor
        & 87.76 & 83.95 & 80.41
        & 85.71 & 80.14 & 73.88
        & 86.73 & 82.04 & 77.14
        & 86.67 & 71.11 & 60.00 \\
        CAP-Evolve
        & 96.33 & 93.61 & 89.39
        & 88.16 & 85.17 & 80.41
        & 92.24 & 89.39 & 84.90
        & 93.33 & 84.44 & 73.33 \\
        \rowcolor{gray!12}
        \tech
        & \textbf{97.55} & \textbf{95.24} & \textbf{92.65}
        & \textbf{93.47} & \textbf{89.25} & \textbf{84.08}
        & \textbf{95.51} & \textbf{92.24} & \textbf{88.37}
        & \textbf{93.33} & \textbf{93.33} & \textbf{93.33} \\
        \bottomrule
    \end{tabular*}
\end{table}

\noindent\textbf{Effectiveness.}
As shown in the last row of Table~\ref{tab:main-results}, \tech achieves 95.51\% (=468/490) pass@3 and 88.37\% (=433/490) pass$^3$ on the 490 SWE-Skills-Bench tasks.
Across the 1,470 evaluation runs on these tasks, it achieves 92.24\% (=1356/1470) trial-level accuracy.
On CannBot, it succeeds on 14 of 15 tasks in every trial, yielding 93.33\% on all three metrics.
These results indicate that \tech can evolve skills that enable the code agent to achieve high success rates on held-out tasks and execute reliably across repeated trials.

\noindent\textbf{Efficiency.}
We examine the average time spent in an evolution loop.
Each loop involves task execution by the code agent, as well as analysis and revision by the \tech framework.
Specifically, the latter includes trajectory abstraction, suspicious actions localization, and skill edit-site localization and revision.
\tech takes approximately 49 minutes per loop for a SWE-Skills-Bench skill and 127 minutes for CannBot.
In particular, code-agent task execution takes 19.8 and 97.8 minutes, respectively.
The three framework components take 3.1, 7.4, and 18.3 minutes on SWE-Skills-Bench, respectively, while the corresponding times on CannBot are 4.4, 5.6, and 19.0 minutes.
Although code-agent execution takes nearly five times as long on CannBot as on SWE-Skills-Bench, the combined time spent on framework analysis and revision remains similar across the two benchmarks.
Consequently, code-agent execution accounts for most of the difference in evolution time between the two benchmarks.

\noindent\textbf{Generalizability.}
\tech remains effective not only across different task splits within SWE-Skills-Bench but also across skill systems that differ in structure, scale, and task domain.
Under the two complementary task splits described in Section~\ref{sec:experimental-setup}, A\(\rightarrow\)B achieves 97.55\% (=239/245) pass@3, 95.24\% (=700/735) trial-level accuracy, and 92.65\% (=227/245) pass$^3$, while B\(\rightarrow\)A achieves 93.47\% (=229/245), 89.25\% (=656/735), and 84.08\% (=206/245), respectively.
Although performance differs between the two split directions, both achieve high held-out scores across all three metrics, indicating that the evolved skills generalize to tasks unseen during evolution.
SWE-Skills-Bench evaluates compact, independent skills for repository-level software engineering, whereas CannBot involves larger, coordinated skills for AscendC kernel generation.
The high scores across all three metrics on CannBot show that the same evolution workflow remains effective with more extensive guidance distributed across multiple files and skills.

\noindent\textbf{Answer to RQ1.}
\tech achieves high held-out performance and reliable execution under both SWE-Skills-Bench split directions and on CannBot's larger, coordinated skill system.
Its framework processing time remains comparable across the two benchmarks, with CannBot's longer evolution time mainly attributable to code-agent task execution.

\subsection{RQ2: Comparison against Baselines}
\label{sec:rq2}

\noindent\textbf{Overall Effectiveness Comparison.}
Table~\ref{tab:main-results} also compares the effectiveness of \tech with that of the original skills and four existing skill-evolution methods.
The results show that \tech achieves the highest pass@3, trial-level accuracy, and pass$^3$ on both benchmarks.

Compared with the original skills, \tech increases pass@3 from 82.45\% to 95.51\% on SWE-Skills-Bench and from 80.00\% to 93.33\% on CannBot, corresponding to relative improvements of 15.84\% and 16.67\%, respectively.
For trial-level accuracy, relative gains of 22.27\% and 40.00\% raise the scores from 75.44\% to 92.24\% and from 66.67\% to 93.33\%, respectively.
More notably, pass$^3$ increases from 67.96\% to 88.37\% on SWE-Skills-Bench, a 30.03\% relative gain, and doubles from 46.67\% to 93.33\% on CannBot.
These results show that all three metrics improve, with the largest relative gains in pass$^3$ indicating more consistent success with the evolved skills.

Against existing methods on SWE-Skills-Bench, \tech solves 16 more tasks than CAP-Evolve, the strongest baseline, and 48 more than EvoSkill, corresponding to pass@3 gains of 3.27 and 9.80 percentage points.
More notably, the gain is larger for consistently solved tasks in every baseline comparison, ranging from 17 to 68 tasks (3.47--13.88 points in pass$^3$).
Consistent with these gains, \tech records 42--174 additional successful trials out of 1,470, raising trial-level accuracy by 2.86--11.84 percentage points.
The advantage in execution consistency is particularly clear on CannBot.
Although CAP-Evolve matches \tech in pass@3 (14/15), \tech solves all 14 tasks in every trial, compared with 8--11 among the four baselines, giving a pass$^3$ advantage of 20.00--40.00 percentage points.
This greater consistency is also reflected at the trial level, where \tech records 4--10 additional successes out of 45 trials and raises accuracy by 8.89--22.22 percentage points.
These results indicate that \tech's advantage extends from covering additional tasks to supporting more consistent success across repeated executions.

\begin{figure}[!tb]
\centering
\begingroup%
\providecommand{\tech}{SkillMorph}%
\definecolor{fgImproved}{HTML}{447E9B}%
\definecolor{fgRegressed}{HTML}{B86558}%
\definecolor{fgUnchanged}{HTML}{DEE2E5}%
\definecolor{fgGrid}{HTML}{E9E9E9}%
\definecolor{fgAxis}{HTML}{BFBFBF}%
\definecolor{fgInk}{HTML}{242424}%
\def\fgFont{\sffamily\fontsize{7.5}{8.8}\selectfont}%
\def\fgTitleFont{\sffamily\fontsize{7.7}{9}\selectfont}%
\ifdefined\fineGrainedFigureBox\else\newsavebox{\fineGrainedFigureBox}\fi%
\begin{lrbox}{\fineGrainedFigureBox}%
\begin{tikzpicture}[x=1cm,y=0.55cm,font=\fgFont,text=fgInk]%
    \pgfmathsetmacro{\aLeft}{2.05}%
    \pgfmathsetmacro{\aRight}{5.22}%
    \pgfmathsetmacro{\aScale}{(\aRight-\aLeft)/240}%
    \pgfmathsetmacro{\aZero}{\aLeft+75*\aScale}%
    \pgfmathsetmacro{\bLeft}{5.85}%
    \pgfmathsetmacro{\bRight}{8.57}%
    \pgfmathsetmacro{\bScale}{(\bRight-\bLeft)/15}%
    \pgfmathsetmacro{\bZero}{\bLeft+6*\bScale}%
    \pgfmathsetmacro{\cLeft}{9.18}%
    \pgfmathsetmacro{\cRight}{13.65}%
    \pgfmathsetmacro{\cScale}{(\cRight-\cLeft)/98}%

    \foreach \x/\fgLegendColor/\fgLegendLabel in {%
        3.00/fgImproved/Improved,%
        6.40/fgUnchanged/Unchanged,%
        9.70/fgRegressed/Regressed} {%
        \fill[\fgLegendColor] (\x,6.04) rectangle ({\x+0.33},6.38);%
        \node[anchor=west] at ({\x+0.44},6.21) {\fgLegendLabel};%
    }%
    \node[anchor=west,font=\fgTitleFont] at (\aLeft,5.22) {(a) SWE tasks};%
    \node[anchor=west,font=\fgTitleFont] at (\bLeft,5.22) {(b) CannBot tasks};%
    \node[anchor=west,font=\fgTitleFont] at (\cLeft,5.22) {(c) SWE skill-level changes};%

    \foreach \tick in {-50,0,50,100,150} {%
        \pgfmathsetmacro{\x}{\aZero+\tick*\aScale}%
        \draw[fgGrid,line width=0.3pt] (\x,-0.52)--(\x,4.60);%
        \draw[fgAxis,line width=0.4pt] (\x,-0.52)--(\x,-0.66);%
        \pgfmathtruncatemacro{\tickLabel}{abs(\tick)}%
        \node[anchor=north] at (\x,-0.79) {\tickLabel};%
    }%
    \foreach \tick in {-4,0,4,8} {%
        \pgfmathsetmacro{\x}{\bZero+\tick*\bScale}%
        \draw[fgGrid,line width=0.3pt] (\x,-0.52)--(\x,4.60);%
        \draw[fgAxis,line width=0.4pt] (\x,-0.52)--(\x,-0.66);%
        \pgfmathtruncatemacro{\tickLabel}{abs(\tick)}%
        \node[anchor=north] at (\x,-0.79) {\tickLabel};%
    }%
    \foreach \tick in {0,25,50,75,98} {%
        \pgfmathsetmacro{\x}{\cLeft+\tick*\cScale}%
        \draw[fgGrid,line width=0.3pt] (\x,-0.52)--(\x,4.60);%
        \draw[fgAxis,line width=0.4pt] (\x,-0.52)--(\x,-0.66);%
        \pgfmathtruncatemacro{\tickLabel}{abs(\tick)}%
        \node[anchor=north] at (\x,-0.79) {\tickLabel};%
    }%
    \foreach \fgLeft/\fgRight in {\aLeft/\aRight,\bLeft/\bRight,\cLeft/\cRight} {%
        \draw[fgAxis,line width=0.4pt] (\fgLeft,-0.52)--(\fgRight,-0.52);%
    }%

    \foreach \method/\row/\sweImproved/\sweRegressed/\cannImproved/\cannRegressed/\improved/\unchanged/\regressed in {%
        SkillOpt/4/109/49/4/4/49/30/19,%
        EvoSkill/3/82/43/5/4/35/47/16,%
        SkillAdaptor/2/75/26/4/1/31/56/11,%
        CAP-Evolve/1/122/36/7/1/54/28/16,%
        {\tech}/0/126/9/7/0/60/33/5} {%
        \ifnum\row=0\def\fgWeight{\bfseries}\else\def\fgWeight{\mdseries}\fi%
        \node[anchor=east,font=\fgFont\fgWeight] at (1.72,\row) {\method};%

        \pgfmathsetmacro{\positiveA}{\aZero+\sweImproved*\aScale}%
        \pgfmathsetmacro{\negativeA}{\aZero-\sweRegressed*\aScale}%
        \fill[fgImproved] (\aZero,{\row-0.23}) rectangle (\positiveA,{\row+0.23});%
        \fill[fgRegressed] (\negativeA,{\row-0.23}) rectangle (\aZero,{\row+0.23});%
        \node[anchor=west,font=\fgFont\fgWeight] at ({\positiveA+0.06},\row) {\sweImproved};%
        \node[anchor=east,font=\fgFont\fgWeight] at ({\negativeA-0.06},\row) {\sweRegressed};%

        \pgfmathsetmacro{\positiveB}{\bZero+\cannImproved*\bScale}%
        \pgfmathsetmacro{\negativeB}{\bZero-\cannRegressed*\bScale}%
        \fill[fgImproved] (\bZero,{\row-0.23}) rectangle (\positiveB,{\row+0.23});%
        \fill[fgRegressed] (\negativeB,{\row-0.23}) rectangle (\bZero,{\row+0.23});%
        \node[anchor=west,font=\fgFont\fgWeight] at ({\positiveB+0.06},\row) {\cannImproved};%
        \ifnum\cannRegressed=0%
            \node[anchor=east,font=\fgFont\fgWeight] at ({\negativeB-0.06},\row) {0};%
        \else%
            \node[anchor=east,font=\fgFont\fgWeight] at ({\negativeB-0.06},\row) {\cannRegressed};%
        \fi%

        \pgfmathsetmacro{\firstBoundary}{\cLeft+\improved*\cScale}%
        \pgfmathsetmacro{\secondBoundary}{\firstBoundary+\unchanged*\cScale}%
        \fill[fgImproved] (\cLeft,{\row-0.23}) rectangle (\firstBoundary,{\row+0.23});%
        \fill[fgUnchanged] (\firstBoundary,{\row-0.23}) rectangle (\secondBoundary,{\row+0.23});%
        \fill[fgRegressed] (\secondBoundary,{\row-0.23}) rectangle (\cRight,{\row+0.23});%
        \node[text=white,font=\fgFont\fgWeight,inner sep=0pt] at ({(\cLeft+\firstBoundary)/2},\row) {\improved};%
        \node[font=\fgFont\fgWeight,inner sep=0pt] at ({(\firstBoundary+\secondBoundary)/2},\row) {\unchanged};%
        \node[text=white,font=\fgFont\fgWeight,inner sep=0pt] at ({(\secondBoundary+\cRight)/2},\row) {\regressed};%
    }%
    \draw[black!60,line width=0.5pt] (\aZero,-0.52)--(\aZero,4.60);%
    \draw[black!60,line width=0.5pt] (\bZero,-0.52)--(\bZero,4.60);%
    \node at ({(\aLeft+\aRight)/2},-1.98) {Number of tasks};%
    \node at ({(\bLeft+\bRight)/2},-1.98) {Number of tasks};%
    \node at ({(\cLeft+\cRight)/2},-1.98) {Number of skill instances};%
\end{tikzpicture}%
\end{lrbox}%
\ifdim\linewidth<13cm%
  \resizebox{\linewidth}{!}{\usebox{\fineGrainedFigureBox}}%
\else%
  \resizebox{13cm}{!}{\usebox{\fineGrainedFigureBox}}%
\fi%
\endgroup%

\caption{Fine-grained performance changes relative to the original skills.}
\label{fig:evolution-changes}
\ifdefined\Description
\Description{Three panels compare five evolution methods using improved and regressed task counts on SWE-Skills-Bench and CannBot, and improved, unchanged, and regressed skill-instance counts on SWE-Skills-Bench.}
\fi
\end{figure}
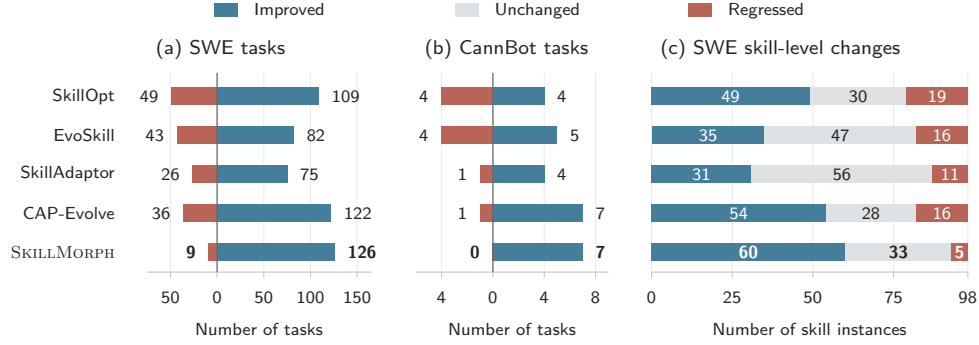

\noindent\textbf{Fine-Grained Comparison.}
We further compare improvements and regressions at the task and skill levels relative to the original skills to understand how the overall gains are distributed.
At the task level, we classify a task as improved, unchanged, or regressed according to whether its number of successful trials out of three increases, remains the same, or decreases.
At the skill level, we use the same categories, comparing pass@3 first and using trial-level accuracy to break ties.
Figure~\ref{fig:evolution-changes} presents task-level changes on both benchmarks and skill-level changes on SWE-Skills-Bench.

On SWE-Skills-Bench, \tech improves a similar number of tasks to CAP-Evolve (126 versus 122), but introduces substantially fewer regressions (9 versus 36).
Specifically, among the 333 originally consistently solved tasks, \tech retains consistent success on 327 (98.20\%), compared with 309 (92.79\%) under CAP-Evolve.
This advantage extends to the other three baselines, which improve 75--109 tasks and regress on 26--49.
These comparisons indicate that \tech's advantage reflects improvements on weaker tasks and better preservation of reliable behavior.

On CannBot, \tech improves task performance while better preserving existing reliable behavior.
Across the 15 tasks, it improves 7 without regression, whereas the four baselines improve 4--7 and regress on 1--4.
Specifically, all 5 tasks initially solved inconsistently now succeed in every trial under \tech, compared with 3--4 under the baselines.
\tech also retains consistent success on all 7 originally consistently solved tasks, whereas the baselines retain only 4--6.
Together, these results highlight \tech's advantage in improving execution consistency while preserving existing capabilities in this more complex, coordinated skill setting.

Beyond individual tasks, we examine how these changes are distributed across the 98 skill instances formed by 49 SWE-Skills-Bench skills under the two complementary task splits.
\tech improves 60 instances, leaves 33 unchanged, and regresses on 5, whereas the four baselines improve 31--54 instances and regress on 11--19.
Thus, the advantage observed at the task level also extends across more skill instances, with fewer regressions than under the existing methods.

\noindent\textbf{Execution Time Comparison.}
We further compare the average agent execution time to examine whether the effectiveness gains come with increased execution costs.
On SWE-Skills-Bench, the average time remains nearly unchanged, decreasing from 16.45 minutes with the original skills to 16.35 minutes with \tech.
On CannBot, however, \tech reduces the average execution time from 141.55 to 87.07 minutes, a reduction of 38.49\%, compared with 110.19--143.26 minutes under the four baselines.
These results suggest that skill evolution can improve execution reliability without increasing average execution time, while also yielding substantial time savings in CannBot's more complex, coordinated skill system.

\noindent\textbf{Answer to RQ2.}
\tech outperforms the original skills and four existing skill-evolution methods in overall effectiveness, improving more skill instances while introducing fewer regressions.
These gains come with nearly unchanged execution time on SWE-Skills-Bench and substantial time savings on CannBot.

\subsection{RQ3: Configurations and Components of \tech}
\label{sec:rq3}
In this section, we evaluate how the number of evolution loops affects \tech's effectiveness and conduct an ablation study to assess the contributions of its key components.

\begin{figure}[t]
    \centering
    \pgfplotslegendfromname{loop-effect-legend}
    \par\smallskip
    \begin{subfigure}[t]{0.48\linewidth}
        \centering
        \begin{tikzpicture}
        \begin{axis}[
            width=0.95\linewidth,
            height=4.0cm,
            xmin=-0.15,xmax=3.15,
            ymin=64,ymax=102,
            xtick={0,1,2,3},
            xticklabels={Original,1,2,3},
            ytick={70,80,90,100},
            ylabel={Success rate (\%)},
            ylabel style={yshift=-10pt},
            axis lines=left,
            axis line style={gray!70,-},
            tick style={gray!70},
            tick label style={font=\scriptsize},
            label style={font=\scriptsize},
            ymajorgrids=true,
            grid style={gray!20,dashed},
            legend to name=loop-effect-legend,
            legend columns=3,
            legend cell align=left,
            legend style={draw=none,font=\scriptsize,column sep=8pt},
        ]
            \addplot[thick,blue!70!black,mark=*,mark size=1.6pt]
                table[x=loop,y expr=100*\thisrow{solved}/490] {figures/loop-effect-swe.dat};
            \addlegendentry{Pass@3}
            \addplot[thick,orange!85!black,dashed,mark=square,mark size=1.9pt,
                mark options={solid,fill=white,line width=0.65pt}]
                table[x=loop,y expr=100*\thisrow{correct}/1470] {figures/loop-effect-swe.dat};
            \addlegendentry{Trial acc.}
            \addplot[thick,green!55!black,densely dotted,mark=triangle,mark size=2.2pt,
                mark options={solid,fill=white,line width=0.65pt}]
                table[x=loop,y expr=100*\thisrow{all}/490] {figures/loop-effect-swe.dat};
            \addlegendentry{Pass$^3$}
        \end{axis}
\end{tikzpicture}
        \captionsetup{margin={29bp,0bp},oneside,
            justification=centering,singlelinecheck=false}
        \caption{SWE-Skills-Bench}
        \label{fig:loop-effect-swe}
    \end{subfigure}
    \hfill
    \begin{subfigure}[t]{0.48\linewidth}
        \centering
        \begin{tikzpicture}
        \begin{axis}[
            width=0.95\linewidth,
            height=4.0cm,
            xmin=-0.15,xmax=5.15,
            ymin=40,ymax=102,
            xtick={0,1,2,3,4,5},
            xticklabels={Original,1,2,3,4,5},
            ytick={40,60,80,100},
            axis lines=left,
            axis line style={gray!70,-},
            tick style={gray!70},
            tick label style={font=\scriptsize},
            label style={font=\scriptsize},
            ymajorgrids=true,
            grid style={gray!20,dashed},
        ]
            \addplot[thick,blue!70!black,mark=*,mark size=1.6pt]
                table[x=loop,y expr=100*\thisrow{solved}/15] {figures/loop-effect-cannbot.dat};
            \addplot[thick,orange!85!black,dashed,mark=square,mark size=1.9pt,
                mark options={solid,fill=white,line width=0.65pt}]
                table[x=loop,y expr=100*\thisrow{correct}/45] {figures/loop-effect-cannbot.dat};
            \addplot[thick,green!55!black,densely dotted,mark=triangle,mark size=2.2pt,
                mark options={solid,fill=white,line width=0.65pt}]
                table[x=loop,y expr=100*\thisrow{all}/15] {figures/loop-effect-cannbot.dat};
        \end{axis}
\end{tikzpicture}
        \captionsetup{margin={15.7bp,0bp},oneside,
            justification=centering,singlelinecheck=false}
        \caption{CannBot}
        \label{fig:loop-effect-cannbot}
    \end{subfigure}
    \caption{Effect of the Number of Evolution Loops on \tech's Performance.}
    \label{fig:loop-effect-comparison}
\end{figure}
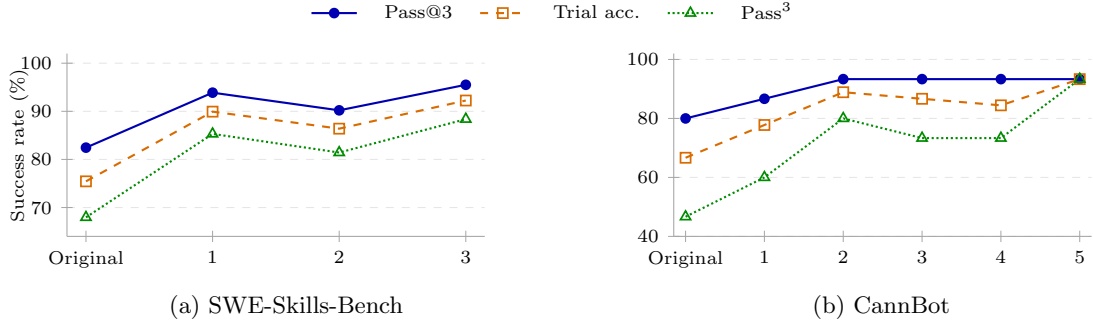

\noindent\textbf{Effect of Evolution Loops.}
We examine the original skills and skills produced after each of the three SWE-Skills-Bench loops and five CannBot loops.
Each version is evaluated on the same held-out tasks under the protocol described in Section~\ref{sec:experimental-setup}.
As shown in Figure~\ref{fig:loop-effect-comparison}, the first loop produces large gains in all three metrics on both benchmarks.
Although some scores decline in intermediate loops, all three metrics reach their highest values in the final loop.
On SWE-Skills-Bench, the first loop raises pass$^3$ from 67.96\% to 85.31\% and accounts for 85--88\% of the final gain across the three metrics.
From the first to the third loop, pass$^3$ increases by 3.06 percentage points, compared with 1.63 points for pass@3, despite a decline in the second loop.
On CannBot, the first loop increases pass@3 from 12 to 13 of 15 tasks and pass$^3$ from 7 to 9.
Pass@3 reaches 14 tasks by the second loop and remains there, whereas pass$^3$ falls from 12 to 11 in the intermediate loops before reaching 14 in the fifth.
Together, these results show that early loops produce most of the gains in task coverage, while later loops can improve execution consistency despite intermediate declines.

\begin{table}[tb]
    \centering
    \caption{Performance of \tech variants with different component configurations (\%).}
    \label{tab:component-ablation}
    \small
    \setlength{\tabcolsep}{5pt}
    \renewcommand{\arraystretch}{1.08}
    \begin{tabular*}{\linewidth}{@{\extracolsep{\fill}}lcccccc@{}}
        \toprule
        \multirow{2}{*}{Variant}
        & \multicolumn{3}{c}{SWE Overall}
        & \multicolumn{3}{c}{CannBot} \\
        \cmidrule(lr){2-4}\cmidrule(l){5-7}
        & Pass@3 & Trial & Pass$^3$ & Pass@3 & Trial & Pass$^3$ \\
        \midrule
        Single-run & 90.20 & 87.69 & 84.90 & 80.00 & 77.78 & 73.33 \\
        w/o History & 93.88 & 89.93 & 85.31 & 93.33 & 88.89 & 80.00 \\
        w/o Localization & 91.22 & 87.35 & 82.24 & 86.67 & 75.56 & 60.00 \\
        \rowcolor{gray!12}
        Full \tech & \textbf{95.51} & \textbf{92.24} & \textbf{88.37} & \textbf{93.33} & \textbf{93.33} & \textbf{93.33} \\
        \bottomrule
    \end{tabular*}
\end{table}

\noindent\textbf{Effect of Components.}
We conduct an ablation study to assess the contributions of repeated execution, the use of history, and skill edit-site localization.
We construct three variants of \tech:
\emph{Single-run} executes each evolution task once per loop (\(m=1\)) while retaining the same number of loops;
\emph{w/o History} omits the historical adjustment of action coverage counts and excludes historical information from proposal generation and evaluation;
and \emph{w/o Localization} skips Skill Edit-Site Localization and generates revisions directly from the selected suspicious actions and their execution contexts.
Table~\ref{tab:component-ablation} reports the performance of \tech variants with different component configurations on both benchmarks, measured by pass@3, trial-level accuracy, and pass$^3$.
Single-run has the largest effect on pass@3, lowering it by 5.31 (=26/490) and 13.33 (=2/15) percentage points on SWE-Skills-Bench and CannBot, respectively.
Removing localization has the largest effect on pass$^3$, lowering it by 6.12 (=30/490) and 33.33 (=5/15) points, respectively.
The w/o History variant has the smallest effect on both benchmarks, with CannBot pass@3 unchanged but lower trial-level accuracy and pass$^3$.
These results suggest that repeated execution contributes most clearly to task coverage, while explicit localization is particularly important for consistent success.

\noindent\textbf{Answer to RQ3.}
(1) The first loop yields large gains, while later loops improve execution consistency, particularly after CannBot's pass@3 levels off.
(2) Single-run has the largest effect on pass@3, w/o Localization on pass$^3$, and w/o History the smallest effect.

\section{Practical Application}
We apply \tech in collaboration with a team that develops AI operators to evolve the skills used in automated kernel generation.
The evaluation covers 63 AI operators of varying complexity, with test cases spanning multiple data types and tensor shapes.
Execution evidence from these tasks guides skill revisions addressing numerical correctness, kernel performance, and the AI operators' development and verification workflow.
We have submitted these skill revisions in 6 pull requests, all of which have been accepted by the developers.
The collaboration is ongoing, and the team continues to use and refine the revised skills.

For tasks that pass correctness checks with both the original and evolved skills, the average completion time decreases from 123.63 to 72.81 minutes, a reduction of 41.11\%.
These savings show that refining workflow guidance to reduce unnecessary operations, such as unfocused searches, can improve efficiency even for tasks that the original skills already solve.

For kernels that pass correctness checks, we measure speedup over the official reference implementation using device-side execution times.
Across 1,788 case-level measurements, 64.26\% and 54.14\% achieve speedups of at least \(0.6\times\) and \(0.8\times\), respectively.
In 44.18\% of these cases, the generated kernels match or exceed the reference performance.

\section{Discussion}

\begin{figure}[tb]
    \centering
    \begin{minipage}{0.6\linewidth}
    \centering
    \begin{tikzpicture}
        \begin{axis}[
            width=0.95\linewidth,
            height=3.6cm,
            xmin=-0.2,xmax=5.2,
            ymin=55,ymax=100,
            xtick={0,1,2,3,4,5},
            xticklabels={0,1,5,10,15,20},
            ytick={60,70,80,90,100},
            xlabel={Number of selected actions (\(n\))},
            ylabel={Success rate (\%)},
            axis lines=left,
            axis line style={gray!70},
            tick style={gray!70},
            tick label style={font=\scriptsize},
            label style={font=\scriptsize},
            ylabel style={yshift=-10pt},
            xlabel style={yshift=2pt},
            ymajorgrids=true,
            grid style={gray!20,dashed},
            legend columns=3,
            legend cell align=left,
            legend style={
                at={(0.5,1.01)},
                anchor=south,
                draw=none,
                font=\scriptsize,
                column sep=3pt,
                inner sep=1pt
            },
        ]
            \addplot[
                thick,blue!70!black,
                mark=*,mark size=1.4pt
            ]
                table[x=position,y expr=100*\thisrow{solved}/490] {figures/top-n-effect.dat};
            \addlegendentry{Pass@3}

            \addplot[
                thick,orange!85!black,dashed,
                mark=square,mark size=1.6pt,
                mark options={solid,fill=white,line width=0.65pt}
            ]
                table[x=position,y expr=100*\thisrow{correct}/1470] {figures/top-n-effect.dat};
            \addlegendentry{Trial acc.}

            \addplot[
                thick,green!55!black,densely dotted,
                mark=triangle,mark size=1.8pt,
                mark options={solid,fill=white,line width=0.65pt}
            ]
                table[x=position,y expr=100*\thisrow{all}/490] {figures/top-n-effect.dat};
            \addlegendentry{Pass$^3$}
        \end{axis}
    \end{tikzpicture}
    \end{minipage}
    \caption{Effect of top-\(n\) action selection.}
    \label{fig:top-n-effect}
\end{figure}
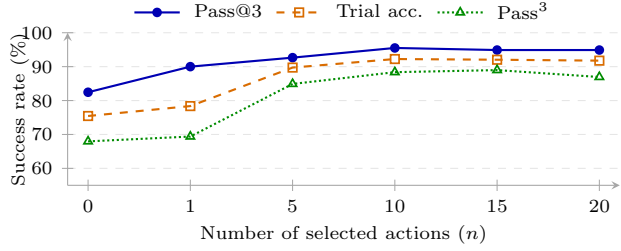

\noindent\textbf{Effect of the Number of Selected Actions}
The preceding experiments show that \tech achieves strong performance when selecting the top-10 suspicious actions in each loop.
To examine how performance depends on this number, we conduct evolution experiments on SWE-Skills-Bench and vary \(n\) among 1, 5, 10, 15, and 20.
All other experimental settings remain unchanged.
Figure~\ref{fig:top-n-effect} presents the results, where \(n=0\) denotes the original skills before evolution.

At \(n=1\), pass@3 reaches 90.00\%, but pass$^3$ is only 69.39\%, indicating high task coverage with limited execution consistency.
As \(n\) increases to 5 and 10, all three metrics improve, with pass$^3$ rising to 84.90\% and 88.37\%, respectively, narrowing the gap with pass@3.
Manual inspection shows that revisions guided by the additional actions address skill deficiencies affecting execution reliability, improving consistency across trials.
Performance remains relatively stable as \(n\) further increases to 15 and 20.
This pattern suggests that selecting too few actions may leave important skill deficiencies affecting execution consistency unaddressed, whereas additional actions beyond a certain selection size yield limited further gains.

\noindent\textbf{Lessons Learned}
Based on \tech, we derive the following implications.

\noindent\(\bullet\)\textit{Skill Authoring.}
We observe that instructions to test thoroughly can still leave verification blocked after the requested code changes are completed.
In these cases, the skill lacks concrete guidance on dependency handling, verification environments, or re-verification after repairs.
This suggests that skill authors should specify the applicability conditions, required execution order, and completion criteria of key operations, using actual trajectories to identify gaps in the guidance.
This observation aligns with prior work emphasizing procedural guidance and verification feedback~\cite{skill-smells}.
When such guidance spans multiple files, the main instructions should clearly reference the relevant materials, and related rules should remain consistent across documents and scripts.

\noindent\(\bullet\)\textit{Task-Specific Adaptation.}
Experience accumulated through cross-task evolution may still leave difficulties specific to the current task unresolved.
In our bidirectional cross-validation, 21 tasks failed in all three evaluation runs with both the original skill and the skill evolved using the opposite task fold.
We examined the evolution records in which these tasks themselves belonged to the task set used for evolution.
Relative to their initial executions in those records, five tasks subsequently succeeded in at least one run, while four others showed improvements in individual verification checks.
These observations suggest that task-specific feedback may complement experience accumulated across tasks.
Reflexion~\cite{reflexion} uses feedback-driven reflection to guide subsequent attempts, while Voyager~\cite{voyager} stores verified programs for reuse.
Building on these mechanisms, future work could explore combining task-level adaptation with \tech to address immediate blockers and turn verified recovery operations and their applicability conditions into reusable skill guidance.

\noindent\textbf{Threats to Validity}
The main threat to \textit{internal} validity arises from stochastic execution and the adaptation of the compared approaches.
To reduce this threat, we use their official implementations under the common experimental protocol described in Section~\ref{sec:experimental-setup} and evaluate each task in three independent trials.
The main threat to \textit{external} validity lies in the selection of skill systems and tasks.
To mitigate this threat, we evaluate \tech on held-out tasks from SWE-Skills-Bench and CannBot, which differ in task domain and skill structure.
We also swap the tasks used for evolution and evaluation on SWE-Skills-Bench to reduce dependence on a particular task split.
The main threat to \textit{construct} validity lies in the metrics used to assess skill quality.
To reduce this threat, we use pass@3, trial-level accuracy, and pass$^3$ to provide complementary measures of task completion and consistency across repeated executions.

\section{Related Work}
\label{sec:related-work}

\noindent\textbf{Agent Skill Construction and Evolution.}
Agent skills package reusable procedural knowledge, while their utility depends on the compatibility among the skill, task, and execution context~\cite{skillsbench,swe-skills-bench}.
Existing methods for improving agent skills can be broadly divided into skill construction and skill evolution.
Skill construction derives new artifacts from execution experience, such as executable libraries, consolidated trajectory lessons, or skills authored by a meta-skill~\cite{voyager,trace2skill,skillevolver}.
Skill evolution instead revises existing artifacts from execution feedback, either through optimization loops based on failure analysis, generated evaluation tasks, or validation-gated edits~\cite{evoskill,skillscoach,skillopt}, or by diagnosing defects from trajectories~\cite{skillrevise,skilladaptor,skillgrad}.
\tech links suspicious-action analysis across tasks and evolution loops to explicit edit-site localization and coordinated skill revisions.

\noindent\textbf{Agent Trajectory Analysis and Utilization.}
Beyond skill evolution, agent execution trajectories have also been widely studied for behavior analysis~\cite{agent-trajectory-study, watson}, failure diagnosis~\cite{mast, who-when, famas}, and context management~\cite{context-as-tool, taco, agentdiet}.
For example, Bouzenia and Pradel~\cite{agent-trajectory-study} characterize the behavioral differences between successful and unsuccessful executions of software engineering agents.
FAMAS~\cite{famas} adapts spectrum-based fault localization to multi-agent systems, contrasting successful and failed executions to attribute failures to the responsible agent actions.
AgentDiet~\cite{agentdiet} removes redundant and expired information to reduce coding-agent inference cost.
Unlike these works, which analyze or manage trajectories without revising persistent artifacts, \tech links behavioral evidence to editable skill artifacts and tracks their revisions across evolution loops.

\noindent\textbf{Feedback-Driven Software Improvement.}
Using execution feedback to improve software has a long history~\cite{autonomic-computing, sbse, sbfl-survey}, among which automated program repair (APR) is a representative line of research~\cite{genprog, capgen, patch-correctness, premm}.
APR leverages tests throughout the repair loop: fault localization contrasts passing and failing executions to rank suspicious program entities~\cite{sbfl-survey}, and candidate patches are searched for and validated against test results, as exemplified by GenProg~\cite{genprog}.
Treating the skills as evolvable software, \tech follows this localize-then-revise paradigm, using spectrum-guided execution feedback to localize and revise deficient skill artifacts across evolution loops.

\section{Conclusion}
In this paper, we propose \tech, a skill-evolution approach for LLM-based agents based on trajectory-guided edit-site localization.
\tech explicitly determines the edit sites, i.e., the skill contents that should be revised, before generating any revision, and grounds this decision in behavioral evidence from execution trajectories.
In particular, \tech abstracts raw trajectories into sequences of state transitions, progressively aggregates evidence across repeated executions, evolution loops, and tasks to identify suspicious actions, and then traces each suspicious action back to edit sites that may span multiple skills, together with a rationale linking the skill content to the action.
Experimental results on SWE-Skills-Bench and CannBot demonstrate the effectiveness of \tech.
Compared with the original skills, \tech achieves relative improvements of 15.84\% in pass@3 and 30.03\% in pass$^3$ on SWE-Skills-Bench, and 16.67\% and 100\% on CannBot, respectively, consistently outperforming the improvements achieved by all the four compared approaches.
Moreover, in a deployment with our collaboration partner, \tech produced six skill-revision pull requests that have been accepted for future AI operator development, further showing its practical value.

\bibliographystyle{plain}
\bibliography{main}
\end{document}